\documentclass[superscriptaddress,pra,twocolumn]{revtex4}

\usepackage[latin1]{inputenc}
\usepackage{graphicx}
\usepackage{amsmath}
\usepackage{amssymb}
\usepackage{dsfont}
\usepackage{graphicx}

\makeatletter
\def\erf{\mathop{\operator@font erf}\nolimits}
\makeatother

\newcommand\be{\begin{equation}}
\newcommand\ee{\end{equation}}

\begin{document}

\title{NOON states in cavities}
\author{D. Rodr\'{\i}guez-M\'endez and H.M. Moya-Cessa} \affiliation{INAOE, Coordinaci\'on de Optica, Apdo. Postal 51 y
216, 72000 Puebla, Pue., Mexico}

\begin{abstract}
We show how NOON states may be generated entangling two cavities
by passing atoms through them. The atoms interact with each cavity
via two-photon resonant transitions. We take advantage of the fact
that depending on the state the atom enter (excite or ground), it
leaves or takes two photons per interaction and leaves the
cavities in a pure state.
\end{abstract}
 \maketitle


\section{Introduction}

A major effort has been directed towards the generation of
nonclassical states of  electromagnetic fields, in which certain
observables exhibit less fluctuations (or noise) than in a
coherent state, whose noise is referred to as the standard quantum
limit (SQL). Nonclassical states that have attracted the greatest
interest include (a) macroscopic quatum superpositions of
quasiclassical coherent states with different mean phases or
amplitudes, also called "Schr\"odinger cats" \cite{1,2,2-5}, (b)
squeezed states \cite{3,3-5} whose fluctuations in one quadrature
or the amplitude are reduced beyond the SQL, (c) the particularly
important limit of extreme squeezing, i.e. Fock or number states
\cite{4} and more recently, (d) nonclassical states of combined
photon pairs also called NOON states \cite{5,6}. It is well known
that these multiphoton entangled states, can be used to obtain
high-precision phase measurements, becoming more and more
advantageous as the number of photons grows. Many applications in
quantum imaging, quantum information and quantum metrology
\cite{7} depend on the availability of entangled photon pairs
because entanglement is a distinctive feature of quantum mechanics
that lies at the core of many new applications. These maximally
path-entangled multiphoton states may be written in the form

\begin{equation}
\left| {N00N} \right\rangle _{a,b}  = \frac{1}{{\sqrt 2 }}\left(
{\left| N \right\rangle _a \left| 0 \right\rangle _b  + \left| 0
\right\rangle _a \left| N \right\rangle _b } \right).
\end{equation}
In the case of cavities \cite{8,9,10,11}, which we  will study in
this communication, this state contains N indistinguishable
photons in an equal superposition of all being in cavity A or
cavity B.

It has been pointed out that NOON states manifest unique coherence
properties by showing that they exhibit a periodic transition
between spatially bunched and antibunched states when   undergo
Bloch oscillations. The period of the bunching/antibunching
oscillation is N times faster than the period of the oscillation
of the photon density \cite{12}.

Most schemes to produce NOON states are in the optical regime
\cite{5,6}.  In this contribution, we would like to analyze the
microwave regime  \cite{8,9,10,11} where  we will show how to
generate NOON states in cavities, by passing atoms through them in
such a way that the cavities get entangled. We consider two-photon
resonant transitions and quality factors sufficiently high such
that it allows us to neglect dissipative effects \cite{4,13,14}.
High quality factor cavities may be constructed in the microwave
regime, with factors of $10^7$ \cite{9} to $10^{10}$ \cite{11}.
Rydberg atoms (usually $^{85}$Rb atoms) with excited state
$40S_{1/2}$ and ground state $39S_{1/2}$, may be passed through
them, either to produce non-classical states and/or to measure
field properties. Here we will show that atoms can entangle two
cavities like the ones shown in Fig. 1, such that NOON states for
microwave quantized stationary fields may be generated
\begin{figure}[hbt]
\begin{center}
\includegraphics[width=0.50\textwidth]{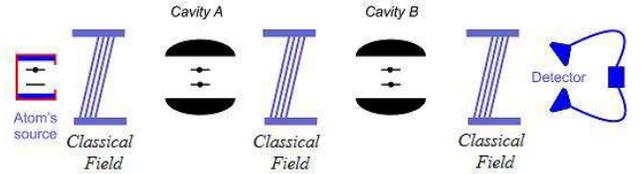}
\end{center}
\caption{\label{fig1} Proposed experimental setup to produce NOON
states in entangled cavities. An excited atom may pass cavities A
and B and leave exactly 2 photons or get entangled with the
cavities. The three classical field are used to prepare the atoms
either in ground, excited or superposition of ground and excited
states.}
\end{figure}

\section{Two-photon dynamics}

In this section we study the atomic behavior when light interacts
with matter in a two-photon resonant transition. We consider the
two-photon interaction Hamiltonian [4,11,14]

\begin{equation}
H_I^{(a)}  = (\Delta+\chi a^{\dagger}a )\sigma_z+\lambda \left(
{a^2 \sigma _ + + a^{ \dagger 2} \sigma _ -  } \right)
\end{equation}

\bigskip
\noindent
 where $\lambda$  is the coupling constant, $a$ and $a^{ \dagger }$
 are the annihilation and creation operators for the field mode (cavity A),
 respectively, $\sigma _ +   = \left| e \right\rangle \left\langle g \right|$ and   $
\sigma _ -   = \left| g \right\rangle \left\langle e \right|$, are
the Pauli spin-flip operators for the two-photon transitions, here
$ \left| e \right\rangle \left (| g \right\rangle)$ means excited
(ground) atomic state. We consider the intermediate state to be so
far from resonance that it can be adiabatically eliminated to give
an effective two-photon coupling of the above form. It contains an
Stark shift tat leads to an intensity dependent transition
frequency. The Stark shift coefficient is denoted as $\chi$ and
$\Delta=\omega_0-2\omega$ is the detuning, where $\omega_0$ is the
unperturbed atomic transition frequency and $\omega$ is the cavity
field frequency. Knight and Shore \cite{shore} have investigated
the validity of the adiabatic elimination for a single atom
evolution.

\bigskip
\noindent The evolution operator is given by (in the atomic basis,
see \cite{Phoenix})
\begin{equation}
U_I^{(a)} \left( t \right) = e^{ - iH_I^{(a)} t}  =e^{i\frac{\chi
t}{2}} \left( {\begin{array}{*{20}c}
   C_{\hat{n}} &  - iS_{\hat{n}}^{\dagger}a^2  \\
   -ia^{ \dagger 2}S_{\hat{n}} &   C_{\hat{n}-2}
\end{array}} \right), \label{evola}
\end{equation}
where
\begin{equation}
C_{\hat{n}}=\cos(\delta_{\hat{n}}t)-i\frac{\Gamma_{\hat{n}}}{\delta_{\hat{n}}}\sin(\delta_{\hat{n}}t),\qquad
S_{\hat{n}}=\lambda\frac{\sin(\delta_{\hat{n}}t)}{\delta_{\hat{n}}},
\end{equation}
with
\begin{equation}
\delta_{\hat{n}}^2=\Gamma_{\hat{n}}^2+\lambda^2(\hat{n}+1)(\hat{n}+2),\qquad
\Gamma_{\hat{n}}=\frac{\Delta+\chi(\hat{n}+1)}{2},
\end{equation}
and $\hat{n}=a^{\dagger}a$.
\begin{figure}[hbt]
\begin{center}
\includegraphics[width=.5\textwidth]{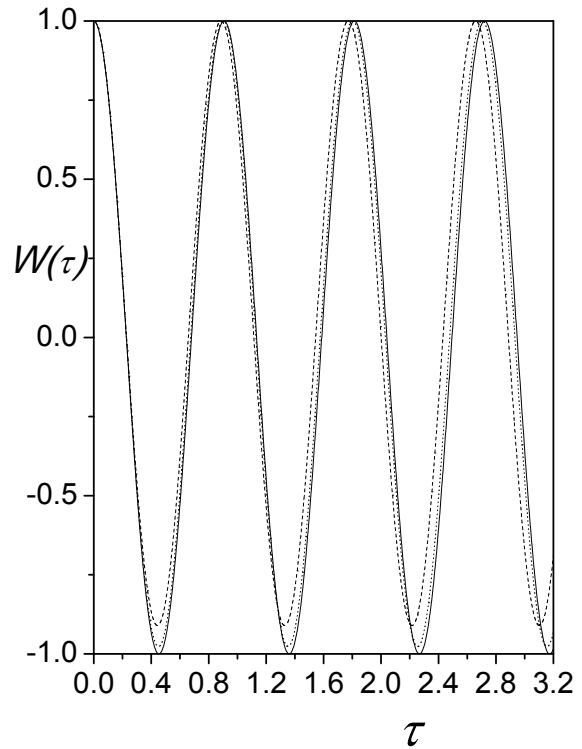}
\end{center}
\caption{\label{fig1} Plot of the atomic inversion, the atom is
initially in its excited state. The field is in the number state
$|2\rangle_a$. Solid curve is for $\chi/\lambda=0$ and
$\Delta/\lambda=0$, dashed curve is for $\chi/\lambda=0.5$ and
$\Delta/\lambda=0$ and dot-dashed curve is $\chi/\lambda=0$ and
$\Delta/\lambda=-0.75$. The scaled time, $\tau$ is defined as
$\tau=\lambda t$.}
\end{figure}
The evolution operator for the second cavity (B), $U_I^{(b)}$,
reads
\begin{equation}
U_I^{(b)} \left( t \right) = e^{ - iH_I^{(a)} t} =e^{i\frac{\chi
t}{2}} \left( {\begin{array}{*{20}c}
   C_{\hat{N}} &  - iS_{\hat{N}}^{\dagger}b^2  \\
   -ib^{ \dagger 2}S_{\hat{N}} &   C_{\hat{N}-2}
\end{array}} \right), \label{evolb}
\end{equation}
with $\hat{N}=b^{\dagger}b$, $b$ the annihilation operator for
cavity B. We  assume that the Stark shift parameters, interaction
constants and detunings are equal for both cavities. In Fig. 2 we
plot the atomic inversion $W(\tau)=P_e(\tau)-P_g(\tau)$, where
$P_e(\tau)$ ($P_e(\tau)$) is the probability to find the atom in
its excited (ground) state, provided it enters  cavity A initially
in the excited state, as a function of the scaled time
$\tau=\lambda t$. The initial state of the field is the number
state $|2\rangle$. It is plotted for different Stark shift
parameters and detunings. The figure shows that, at several times
the atom is in its ground state, or close to it. In the rest of
the paper we will look in particular at the time $\tau_p\approx
3.16$ as it is a time when the field gains two photons.

\noindent In case we consider the atom initially in its ground
state, it may be shown  that the probability to find it in the
ground state at time $ \tau_p $ is approximately zero. In this
case the atom removes, in a clean form, two photons from the
cavity.

\subsection{\emph{Generation of the state} $\left| 2 \right\rangle _a \left| 2 \right\rangle _b$ }
We can generate the state $|2\rangle_a|2\rangle_b$ if we start
from two empty cavities and pass an initially excited atom through
cavity A, let it interact with the vacuum field a time $\tau_p$,
then the atom leaves the cavity in its ground state (See Fig. 2).
After it exits cavity A, a classical field is turn on (second
classical field in Fig. 1), which produces a rotation in the atom
that takes it again to its excited state. Then it passes through
cavity B leaving again two photons in it as it exits (of course
for the same interaction time $\tau_p$). In this way we pass from
the state $|0\rangle_a|0\rangle_b$ to $|2\rangle_a|2\rangle_b$.

\section{NOON states by entangling the cavities}

\noindent We now consider the atom to be in a superposition of its
ground an excited states, i.e.

\begin{equation}
\left| {\psi _{atom} } \right\rangle  = \frac{1}{{\sqrt 2 }}\left(
{\left| e \right\rangle  + \left| g \right\rangle } \right) =
\frac{1}{{\sqrt 2 }}\left( {\begin{array}{*{20}c}
   1  \\
   1  \\
\end{array}} \right).
\end{equation}
 Both cavity fields have been prepared in the state $|2\rangle$. We now
consider the ideal case of $\chi=0$ and $\Delta=0$, after
interaction with cavity A, we obtain (approximately) the
field-atom entangled state
\begin{equation}
\left| {\psi _{a - f} } \right\rangle  = \frac{i}{{\sqrt 2
}}\left( {\begin{array}{*{20}c}
   {\left| 0 \right\rangle _a }  \\
   {\left| 4 \right\rangle _a }  \\
\end{array}} \right),
\end{equation}
\noindent without disturbing the atom, it passes now through the
second cavity, in the number state $|2\rangle_b$, that produces
the field-atom-field entangled state
\begin{equation}
\left| {\psi _{f - a - f} } \right\rangle  =  -
\frac{1}{\sqrt{2}}\left( {\begin{array}{*{20}c}
   {\left| 4 \right\rangle _a \left| 0 \right\rangle _b }  \\
   {\left| 0 \right\rangle _a \left| 4 \right\rangle _b }  \\
\end{array}} \right).
\end{equation}
 When the atom exits cavity B, the last classical field
is turn on, rotating the atom, and therefore producing the state
\begin{equation}
\left| {\psi _{f - a - f} } \right\rangle  =  - \frac{1}{ 2
}\left( {\begin{array}{*{20}c}
   {\left| 4 \right\rangle _a \left| 0 \right\rangle _b  - \left| 0 \right\rangle _a \left| 4 \right\rangle _b }  \\
   {\left| 0 \right\rangle _a \left| 4 \right\rangle _b  + \left| 4 \right\rangle _a \left| 0 \right\rangle _b }  \\
\end{array}} \right).
\end{equation}

\bigskip
\noindent Finally, by detecting the atom in its ground state, the
wave function is collapsed to an entangled states of the two
separate cavities, i.e. to the NOON state

\begin{equation}
\left| {\psi _{f - f} } \right\rangle  = \frac{1}{{\sqrt 2
}}\left( {\left| 0 \right\rangle _a \left| 4 \right\rangle _b  +
\left| 4 \right\rangle _a \left| 0 \right\rangle _b } \right).
\end{equation}

\noindent Of course, detection of the atom in its excited state
would have produced a NOON state with a different sign in between.
\begin{figure}[hbt]
\begin{center}
\includegraphics[width=.5\textwidth]{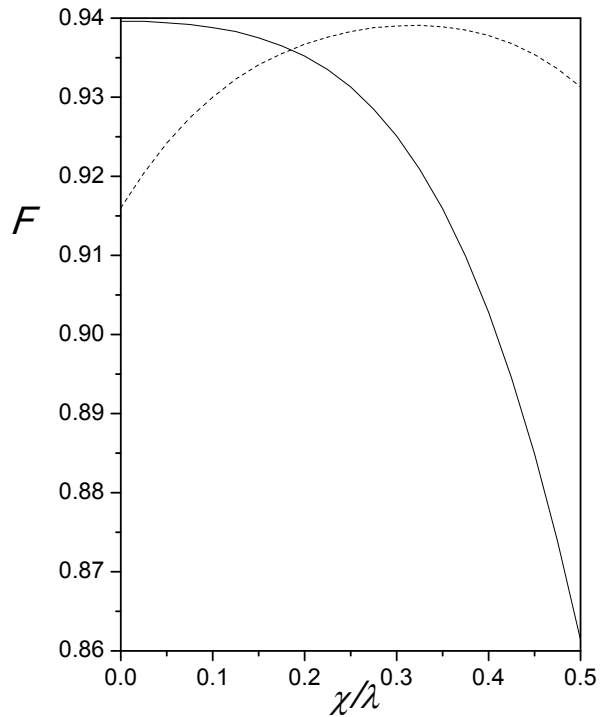}
\end{center}
\caption{\label{fig1} Fidelity as a function of the Stark shift
parameter. Solid line for $\Delta=0$ and dashed line for
$\Delta=-.75\lambda$}
\end{figure}
\subsection{Fidelity}
We now calculate the fidelity  \cite{fidel}
\begin{equation}
F=|_{a,b}\langle|{N00N} | \psi_{f-f}(\tau_p) \rangle|^2,
\end{equation}
which measures the "closeness" of the two quantum states $|
\psi_{f-f}(\tau_p) \rangle$ and $| N00N \rangle_{a,b}$. We find $|
\psi_{f-f}(\tau_p) \rangle$ by applying the evolution operators,
(\ref{evola}) and (\ref{evolb}), to the state
$|2\rangle_a|2\rangle_b\frac{1}{\sqrt{2}}\left(|e\rangle+|b\rangle\right)$.
We plot this function in Fig. 3, where we note a fidelity value $
F\approx 0.94$ for the Stark shift parameter equal to zero. As
this parameter increases the fidelity decreases but remains close
to its initial value. Moreover, it may be increased for non-zero
values of the Stark shift parameter by properly adjusting the
detuning such that it cancels out the effect of the AC Stark
effect. It is clear that dissipative effects would produce the
fidelity also to decrease, as it is a well-known fact that
dissipation erases non-classical features.
\section{Conclusion}
It has been shown that by controlling the interaction time between
of atoms passing through two cavities, cavity fields may be
entangled. A final  detection of the atom in its excited or ground
state yields a NOON state, i.e. and entangled state of both
cavities. In this form we have added another system in which NOON
states may be produced, i.e. extended the regime to microwave
cavities. Measurement of cavity fields may be done via quantum
state reconstruction even in the presence of dissipation
\cite{18}.

\end{document}